# Integrating Uncertainty Quantification into Computational Fluid Dynamics Models of Coronary Arteries Under Steady Flow

**Muhammad Usman**
Department of Biomedical Engineering, Texas A&M University, College Station, TX 77843
2121 W Holcombe Blvd., Alkek Building
Houston, TX 77030
musman1@tamu.edu

**Peter N. Castillo**
Department of Biomedical Engineering, Texas A&M University, College Station, TX 77843
2121 W Holcombe Blvd., Alkek Building
Houston, TX 77030
pnc3717@tamu.edu

**Akil Narayan**
Scientific Computing and Imaging Institute, University of Utah, Salt Lake City, UT
Department of Mathematics, University of Utah, Salt Lake City, UT 84112
72 S Central Campus Drive, Salt Lake City, UT 84112
akil@sci.utah.edu

**Lucas H. Timmins[1]**
Department of Biomedical Engineering, Texas A&M University, College Station, TX 77843
Department of Biomedical Engineering, University of Utah, Salt Lake City, UT 84112
Scientific Computing and Imaging Institute, University of Utah, Salt Lake City, UT 84112
2121 W Holcombe Blvd., Alkek Building
Houston, TX 77030
lucas.timmins@tamu.edu
ASME Fellow

---

[1] Corresponding author

**ABSTRACT**

Computational models are continuously integrated in the clinical space, where they support clinicians in disease diagnosis, prognosis, and prevention strategies. While assisting in clinical space, these computational models frequently use deterministic approaches, where the inherent (aleatoric) variability of input parameters is ignored. This questions the credibility and often hinders the clinical adoption of these computational models. Therefore, in this study, we introduced uncertainty quantification in the computational fluid dynamics models of the left main coronary artery to analyze the influence of input hemodynamics parameters on wall shear stress (WSS). `UncertainSCI` was used, where an emulator was built using polynomial chaos expansion between the input parameters and the output quantity of interest, and the output sensitivities and statistics were directly extracted from the emulator. The uncertainty-informed framework was first applied to an analytical solution of the Navier-Stokes equation (Poiseuille flow) and then to a patient-specific model of the left main coronary artery. Different input hemodynamics parameters are considered, such as pressure, viscosity, density, velocity, and radius, whereas wall shear stress was considered as our output quantity of interest. The results suggest that velocity dominated the variability in WSS in the analytical model (~79%), whereas viscosity dominated in the patient-specific model (~59%). The results further suggest that out of all the Sobol indices interactions, unary interactions were the most dominant ones, contributing ~93.2% and ~99% for the analytical and patient-specific model, respectively. This study will enhance confidence in computational models, facilitating their adoption in the clinical space to improve decision-making for coronary artery disease diagnosis, prognosis, and therapeutic strategies.

*Keywords: coronary artery disease, uncertainty quantification, patient-specific modeling, computational fluid dynamics*

# 1 INTRODUCTION

There is a growing integration of computational fluid dynamics (CFD) modeling techniques into the clinical setting. Modeling results augment anatomic, physiologic, and functional data to advance patient management and inform diagnosis, prognosis, and therapeutic intervention [1]. For example, patient-specific CFD models in cardiovascular medicine have been applied in the setting of congenital heart disease [2], Kawasaki disease [3], pulmonary hypertension [4], abdominal aortic and cerebral aneurysms [5], and coronary atherosclerosis [6], [7]. Unfortunately, most modeling simulations do not account for the intrinsic uncertainties that arise from variability in input parameters (i.e., aleatoric uncertainty) and assumptions made due to limited data (i.e., epistemic uncertainty). In fact, a recent review of *in silico* clinical trials in cardiology evaluating the effectiveness and feasibility of model-directed interventions reported that <20% of studies accounted for input parameter uncertainty [8]. Accordingly, establishing computational frameworks that are robust to model uncertainty is essential to ensure that model-based predictions are accurate, reliable, and safe before their clinical adoption.

Computational uncertainty quantification (UQ) provides an advanced mathematical framework to characterize the variability in simulations due to intrinsic uncertainty in model-dependent quantities of interest [9]. In contrast to deterministic approaches, whereby model input quantities are fixed, and a single simulation provides model results, UQ in the forward model setting formalizes a statistical model for input uncertainty and uses computational algorithms to deliver mathematical and statistical guarantees on output accuracy. Although few studies have integrated UQ into blood flow simulations, analyzing generalized and patient-specific geometries and quantifying sensitivities in flow, pressure [10], and wall shear stress (WSS) [11], [12], [13], [14]. Importantly, however, nearly all prior uncertainty-aware hemodynamic simulations have relied on Monte Carlo (MC) methods for parameter sampling, which can be computationally demanding. As uncertainty-aware CFD frameworks advance toward clinical adoption, the associated computational costs of MC approaches remain unmanageable and impractical. Thus, the integration of efficient sampling strategies is critical to achieving computationally tractable and clinically applicable uncertainty-aware CFD models.

In this study, we leveraged our established UQ-FE framework, which couples the open-source UQ software `UncertainSCI` and the `FEBio` software suite [15] to examine how uncertainty in variables within the Navier-Stokes equations propagates to variability in WSS. The goal of the study was to utilize polynomial chaos expansion (PCE)-based UQ methods to quantify WSS variability derived from analytical or FE approaches under steady flow conditions. Recognizing the clinical interest in understanding the prognostic value of WSS in the natural history of coronary artery disease and plaque vulnerability [16] [17], the focus was directed at blood flow in the epicardial coronary arteries.

## 2 METHODS

The computational framework developed to analyze the influence of uncertainty in input hemodynamic parameters is summarized in Figure 1. Briefly, probability distributions were fit to $m$-input parameters that were sampled to generate $n$-parameter sets (or ensembles). Data sets were integrated into an analytical solution for WSS under steady flow of an incompressible, Newtonian fluid in a rigid cylinder (i.e., Poiseuille flow) or a batch-processing scheme to predict WSS in a patient-specific coronary artery CFD model. Finally, statistical and global sensitivity analyses were carried out to quantify the impact of parameter variability on WSS uncertainty.

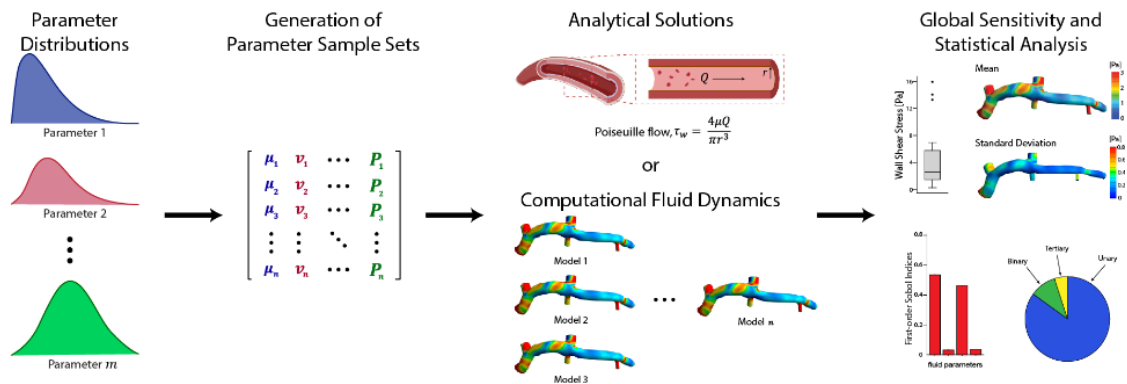

**Fig. 1** Schematic representation of the implemented computational framework to quantify the impact of hemodynamic parameter variability on uncertainty in WSS values derived from an analytical solution and patient-specific FE model.

### 2.1 Hemodynamic Parameter Distributions

The distributions for input parameters were derived from population-based or patient-specific data, as reported in Table 1. Patient-specific distributions were determined by evaluating invasive measurements, as discussed in Section 2.3, across 6 cardiac cycles.

Normal distributions were assumed for density as well as patient-specific velocity and pressure, while gamma distributions were used for viscosity as well as Poiseuille flow velocity and radius to ensure physically meaningful (i.e., positive) values. Distribution parameters (e.g., mean, standard deviation) were specified within `UncertainSCI`, and parameter sets were efficiently sampled across the multi-dimensional parameter space via weighted approximate Fekete points ([18], [19]; Fig. 1).

**Table 1**. Fluid parameter values for analytical and patient-specific model

| Parameter | Poiseuille Flow | Patient-specific |
|---|---|---|
| Viscosity ($\mu$) [mPa·s] | 3.48 ± 0.51 [20] | |
| Velocity ($v$) [m/s] | 0.27 ± 0.16 [21] | 0.145 ± 0.01 |
| Radius ($r$) [mm] | 1.75 ± 0.40 [22] | — |
| Density ($\rho$) [kg/m³] | — | 1051.50 ± 10 [23] |
| Pressure ($P$) [mmHg] | — | 97.54 ± 5.42 |

Data are reported as the mean ± standard deviation. References are provided for population-derived values.

**2.2 Approaches to Quantify Coronary Artery Wall Shear Stress**

**2.2.1 Analytical Solution (Poiseuille Flow)**

The initial implementation of the parameter ensembles utilized an analytical solution to the Navier-Stokes equations. Assuming a steady flow of an incompressible, Newtonian fluid in a rigid cylinder (i.e., Poiseuille flow) [24], WSS is calculated as,

$$\tau_{wall} = \frac{4\mu Q}{\pi r^3}, \tag{1}$$

where $\mu$ is the fluid dynamic viscosity, $Q$ is the flow rate, and $r$ is the cylinder radius. WSS values were determined across parameter sets via a custom `Python` script and stored for UQ analysis.

### 2.2.2 Patient-specific Coronary Artery CFD Model

A representative patient-specific coronary artery geometry, including the left main and left anterior descending coronary artery, was reconstructed utilizing established approaches from data acquired in a prospective study [7], [16]. In brief, biplane angiography and virtual histology intravascular ultrasound (VH-IVUS) images were acquired, as well as Doppler flow and pressure measurements in the interrogated vessel (Fig. 2a). The 3D lumen geometry was reconstructed by stacking the VH-IVUS images perpendicular to the catheter centerline and wrapping a surface to the lumen pointcloud (Fig. 2b). The reconstructed geometry was meshed in `SimVascular` [25] (global edge size = 0.5 mm, 4 boundary layers, decreasing ratio = 0.8), resulting in a mesh size of ~2 million linear tetrahedral elements (Fig. 2c). The volume mesh was imported into `FEBio Studio` [26]. Applied boundary conditions included a steady-state plug velocity profile at the inlet surface and resistance boundary conditions at all outlet surfaces, with values derived from area scaling of the total resistance ($R_{tot}$) of the system, which was determined as,

$$R_{tot} = \frac{P}{Q}, \tag{2}$$

where $P$ and $Q$ are the mean pressure and volumetric flow rate, respectively. A no-slip boundary condition was applied at all wall surfaces. The FE simulations were executed using the CFD solver in `FEBio` [27]. Broyden's quasi-Newton method, in combination with the `mkl_dss` linear solver, was employed to accelerate convergence, and the simulation was executed for 1000 time steps with a step size of 0.001 s. Following solution convergence, data were post-processed to determine WSS across the spatial domain (Fig. 1d). Patient-specific CFD simulations were executed on the `GRACE` and `FASTER` clusters ain the High Performance Research and Computing at Texas A&M University using a SLURM job batch array, with at least 8 models running in parallel, each allocated 8 CPUs (2.2 – 4.0 GHz) and 64GBs of memory.

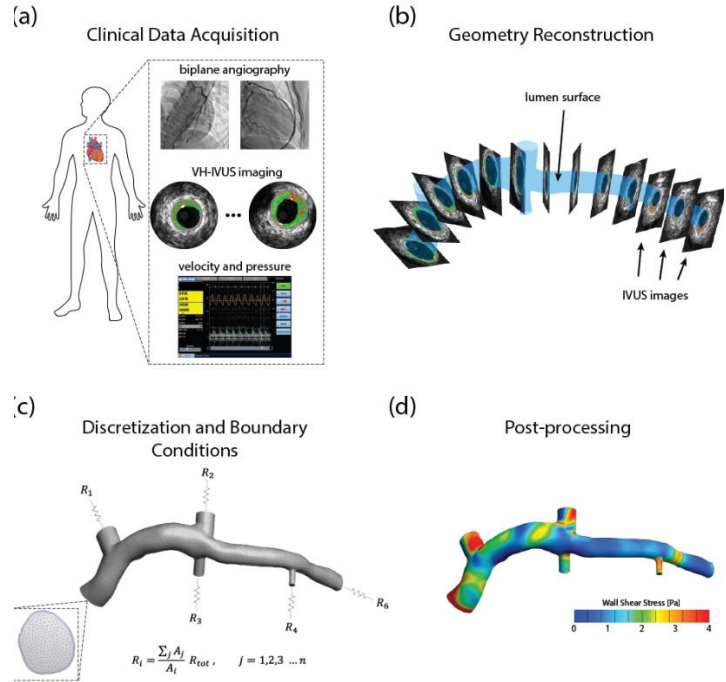

**Fig. 2** Illustration of patient-specific computational fluid dynamics framework, including data acquisition, geometry reconstruction, meshing and boundary conditions, and post-processing.

## 2.3 Uncertainty Quantification Analysis Framework

Extending our previously developed framework [15], we employed the open-source, Python-based software suite, `UncertainSCI` [28], [29], to examine uncertainty in calculated or FE-predicted WSS values. `UncertainSCI` leverages modern PCE techniques to construct emulators, enabling the ability to predict model outputs for general parameter sets and probe model uncertainty. The emulator is a sum of orthogonal, polynomial functions with a maximal order $p$, which is user-specified and results in the generation of $n$-parameter sets via the computationally efficient WAFP procedure [18], [30]. Importantly, the emulator serves as a surrogate model, approximating the mapping between input parameter(s) and model output(s), and enabling the model output to be queried without solving the computationally expensive forward problem. Parameter sets served as inputs for either the analytical solution or an FE batch processing scheme that iterated through ensembles while using the same `FEBio` input file (Fig. 1). Calculated or predicted WSS values were retrieved, permitting emulator construction.

To assess the quality of the PCE (i.e., convergence), the relative error ($\varepsilon_\delta$) was quantified between the emulator approximations and calculated/predicted WSS values, whereby

$$\varepsilon_\delta = \frac{1}{N}\sum_{i=1}^{N}\frac{\|A\hat{x} - \vec{b}\|_2}{\|\vec{b}\|_2}, \qquad (3)$$

where $A\hat{x}$ is the approximation, $\vec{b}$ is the analytical/FE model output, $N$ is the number of elements or takes the value of 1 for the analytical solution, and $\|\cdot\|_2$ indicates the 2-norm of the vector. For orders $p = \{1,2,\ldots 5\}$, $\varepsilon_\delta$ was calculated across 5 independent PCE runs. For the patient-specific model, WSS values for each mesh element were utilized to evaluate the relative error of each element and then averaged spatially and across runs at a given order. In addition, parameter sets were oversampled (2×) in the Poiseuille flow solution to confirm error stability across sampling rates, ensuring aliasing error was minimized (Appendix Fig. A1).

**2.4 Statistical and Global Sensitivity Analyses**

Statistical measures (e.g., mean, standard deviation) were determined directly from the constructed PCE emulator. Also, Sobol's indices [31], which quantify the relative contribution of individual parameters or parameter sets to variance in the output (here, WSS), were computed directly from the polynomial emulator coefficients. The convergence criteria for the PCE emulators were defined as a relative error <0.1% and first-order Sobol index value changes of <0.01 (i.e., index stability) between increasing polynomial orders. Data are reported as the mean $\pm$ standard deviation for normally distributed variables and as median (Q1, Q3) for non-normally distributed variables.

**3 RESULTS**

The number of parameter sets generated across the 5 orders and run times for the analytical solution are shown in Table 2. Parameter sets ranged from 14 to 66 across orders, with a total run time of ~47 seconds for 5 runs at order 5 on a multi-core personal computer (Windows 11 server machine; Intel® Xeon® w5-3433 CPU, 16 cores at 2.0-4.2 GHz). The relative error for WSS between the PCE emulator and analytical solution demonstrated a reduction in magnitude and variability across runs at increasing order (Fig. 3a). For example, error was 5.38 $\pm$ 2.71%, 0.01 $\pm$ 0.004%, and 3.91e-5 $\pm$ 2.1e-5% for orders 1, 3, and 5, respectively. Moreover, first-order Sobol indices stabilized at higher orders (Fig. 3b), with minimal changes in index values (<0.02) and small standard deviations (~0.003) across orders 3 through 5. Similar trends in relative error and Sobol index convergence were observed at a 2× sampling rate (see Appendix). Although order 2 with 2× sampling achieved lower error and Sobol index stabilization (i.e., the PCE emulator converged), it required more samples than order 3 without oversampling, which also met the convergence criteria. Thus, order 3 without oversampling captured the dominant uncertainty modes while remaining computationally efficient.

**Table 2.** Parameter sets and compute time for the Poiseuille flow UQ analysis. Run times are reported for five PCE runs in total at each order, and include parameter set generation, WSS calculation, PCE emulator construction, and statistical and global sensitivity analysis.

| Order | Parameter Sets | Run Time (s) |
|---|---|---|
| 1 | 14 | 13.28 ± 1.31 |
| 2 | 20 | 16.09 ± 1.7 |
| 3 | 30 | 20.99 ± 1.67 |
| 4 | 46 | 37.56 ± 3.39 |
| 5 | 66 | 46.79 ± 5.48 |

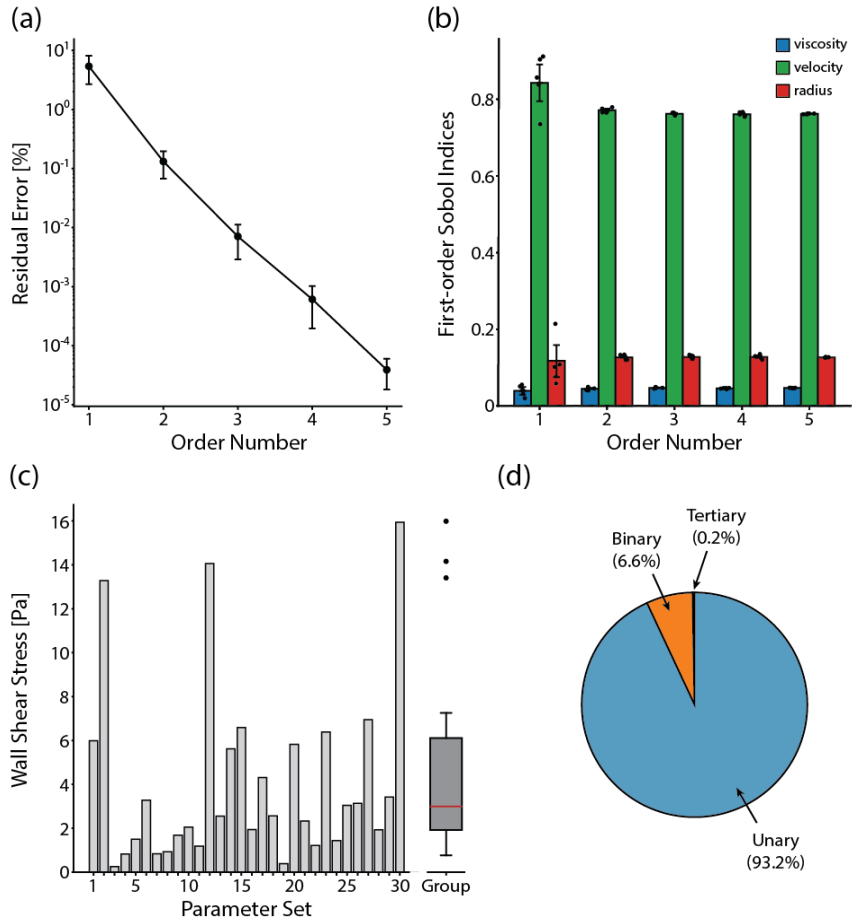

**Fig. 3** PCE emulator and sensitivity indices convergence analysis for the analytical solution – (a) relative error and (b) first-order Sobol indices for each PCE order across 5 runs. (c) Distribution of WSS values from the analytical solution for order 3 (run 1) PCE analysis (solid red line: median, box: interquartile range). (d) Percent output variance due to unary (single parameter), binary (pairwise), and tertiary (triplewise) interactions for order 3 PCE analysis.

Input parameter uncertainties propagated to yield variability in Poiseuille flow-derived WSS values, as evident by the considerable variation across the 30 parameter sets at order 3 (Fig. 3c). Median, minimum, and maximum values were 2.56 (1.45, 5.77), 0.26

($v = 0.051\ \frac{m}{s}$, $\mu = 0.0037\ Pa.s$, $r = 0.002\ m$), and 15.94 ($v = 2.09\ \frac{m}{s}$, $\mu = 0.004\ Pa.s$, $r = 0.002\ m$), respectively. Sensitivity analysis indicated that uncertainty in velocity dominated the variance in WSS. First-order Sobol indices at order 3 were 0.81, 0.13, and 0.05 for velocity, radius, and viscosity, respectively (Fig. 3b). Unary interactions accounted for >93% of the variance in WSS, with binary (i.e., interactions between two parameters) and tertiary interactions accounting for approximately 6.4% and 0.09% of the variance, respectively (Fig. 3d). Across the 3 possible pairwise combinations, the interaction between velocity and radius accounted for 67.4% of variance due to binary interactions; thus, accounting for approximately 4.3% of the total variance (67.4% of 6.4%).

Parameter sets for the patient-specific CFD model ranged from 15 (order 1) to 136 (order 5) and required ~4 to 30 days to run the simulations, respectively (Table 3). As observed in the analytical solution results, the emulator exhibited a reduction in relative error with increasing polynomial order that can be appreciated spatially and quantitatively (Fig. 4a). Error reduced at each element across the computational domain, with average values of 4.27% ± 1.1 %, 0.02 ± 0.002%, and 8e-4 ± 2e-4% at orders 1, 3, and 5, respectively. Furthermore, Sobol indices stabilized at higher orders, again observed spatially and quantitatively (Fig. 4b). Average first-order Sobol indices changed by <0.0008 at order 3 and higher with standard deviations of ~1e-4 across the 5 PCE runs (i.e., order 3 met the defined convergence criteria).

**Table 3.** Patient-specific model parameter sets and compute time across PCE orders. Five PCE runs were performed across each order.

| Order | Parameter Sets | CFD Simulation Run Time (hh:mm:ss) |
|---|---|---|
| 1 | 15 | 90:41:43 ± 2:54:30 |
| 2 | 25 | 108:24:24 ± 28:21:39 |
| 3 | 45 | 240:12:31 ± 6:53:56 |
| 4 | 80 | 399:18:34 ± 64:26:25 |
| 5 | 136 | 627:58:38 ± 5:55:00 |

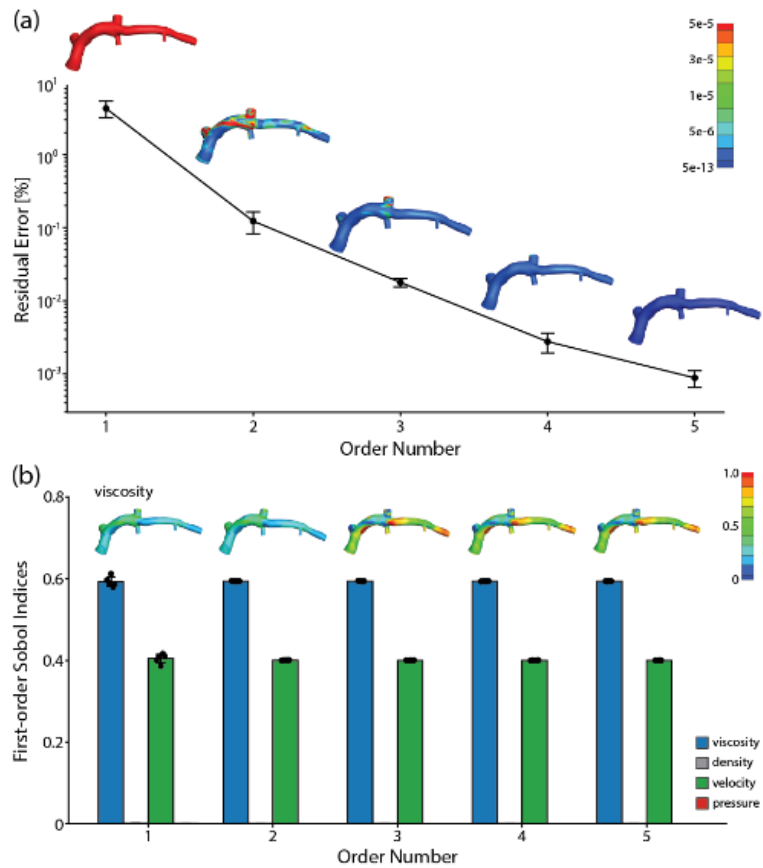

**Fig. 4** PCE emulator and sensitivity indices convergence analysis in the patient-specific CFD model – (a) relative error and (b) first-order Sobol indices for each PCE order across 5 runs. Element-wise data are shown in the fringe plots and summarized in the graphs.

For conciseness, only first-order Sobol indices across the computational domain are presented for viscosity, but the others also demonstrated stability as illustrated in the bar graphs.

Variations in WSS values were observed across the 45 patient-specific CFD models at order 3. Representative models with the highest, average, and lowest mean WSS values are shown in Fig. 5a. Although WSS magnitudes differed at the same spatial locations across the models, the relative spatial distributions within each model were consistent (i.e., areas of elevated and reduced WSS were preserved). The highest, average, and lowest models exhibited median WSS values of 1.33 Pa (0.93, 2.18), 0.91 Pa (0.63, 1.52), and 0.47 Pa (0.31, 0.79), respectively (Fig. 5b). The observed consistency in the relative WSS distributions in space is complemented by the shape of the violin plots.

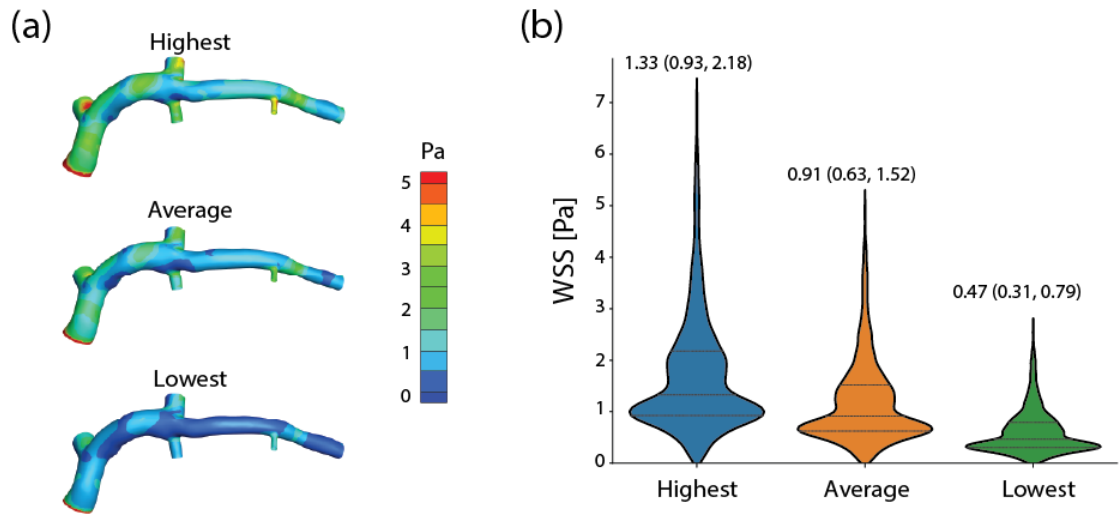

**Fig. 5** Representative WSS (a) fringe plots and (b) violin plots for models exhibiting the highest, average, and lowest mean values at order 3. Violin plots display data between the 10[th] and 90[th] percentiles for visual clarity.

At the converged PCE order of 3 for the CFD models, sensitivity analysis revealed that variability in viscosity and velocity had the greatest impact on WSS variability. Unlike the analytical solution, first-order Sobol indices varied spatially (Fig. 6). The Sobol indices for pressure and density were relatively homogeneous throughout the model, however, those for viscosity and velocity were higher in contrary regions across the model. For example, the Sobol indices for viscosity were high in the inner wall of a branching vessel, whereas velocity values were low (Fig. 6).

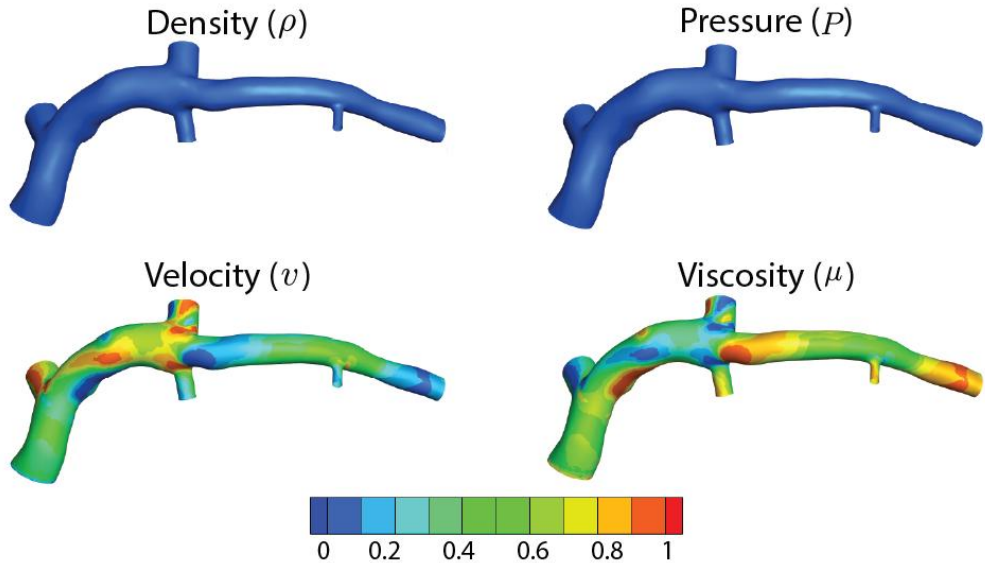

**Fig. 6** Spatial variation in normalized first-order Sobol indices across the patient-specific models at the converged order 3.

Spatially averaged first-order Sobol indices were approximately 0.59, <1e-3, 0.40, and <1e-4 for viscosity, density, velocity, and pressure, respectively (Fig. 7a). Unary interactions were dominant (~99.5%), however, higher-order interactions were present (Fig. 7b). Interactions between two (binary) and three (tertiary) parameters accounted for approximately 0.51% and 0.01%, respectively, of variance in WSS. For the binary interactions, the interaction between velocity and viscosity accounted for ~90% of that variance in WSS (i.e., the interaction between velocity and viscosity accounted for ~0.46% of the total variance in WSS). The interaction between viscosity, velocity, and pressure accounted for ~65% of the variance in WSS due to tertiary interaction.

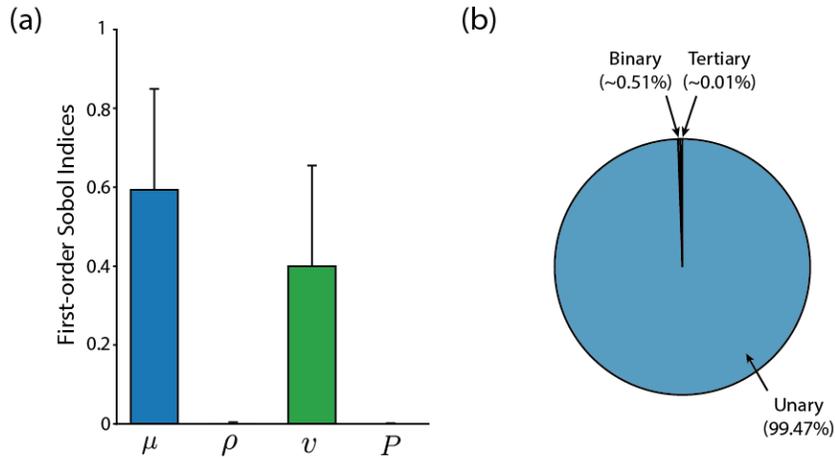

**Fig. 7** Sensitivity analysis at order 3 PCE analysis – (a) normalized first-order Sobol indices for viscosity ($\mu$), density ($\rho$), velocity ($v$), and pressure ($P$), and (b) percent output variance due to unary (single parameter), binary (pairwise combinations), and tertiary (triplewise combinations) interactions.

## 4 DISCUSSION

In this study, we extended our previously developed UQ-FE framework, which coupled `UncertainSCI` with the `FEBio` software suite, and quantified how uncertainty in hemodynamic parameters influences variability in WSS under steady flow conditions derived from an analytical solution and a patient-specific CFD model. The accuracy of the PCE surrogate models was confirmed through convergence testing, demonstrating that a third-order polynomial expansion (i.e., PCE order 3) without parameter set oversampling was sufficiently accurate and provided reliable statistical estimates and stable sensitivity indices. Examining flow in a non-distending cylinder (i.e., Poiseuille flow), we demonstrated that WSS is most sensitive to uncertainties in viscosity and radius, both derived from population data, whereas in a patient-specific left anterior descending coronary artery, values were most sensitive to viscosity (population-based data) and velocity (patient-specific data). Furthermore, the sensitivity analysis highlighted that unary interactions (first-order) were dominant, accounting for >93% of variance in WSS values, indicating limited interaction between input parameters.

UQ has been broadly applied in nearly all biomechanics domains, with applications, for example, in evaluating surgical implants [32], joint kinematics [33], soft tissue mechanics in the brain [34], and myocytes transduction [35]. For blood flow simulations, studies have successfully integrated UQ to quantify the confidence in predicted hemodynamic measures across the vascular tree. Early studies examined parameterized geometries (abdominal aorta aneurysm, carotid bifurcation, stenosed coronary artery) and utilized stochastic collocation sampling methods, enabling efficient exploration of variability in geometry, boundary conditions, and material properties to quantify uncertainties in WSS and fractional flow reserve (FFR) [10], [36]. Results clearly demonstrated the importance of incorporating uncertainties into hemodynamic simulations (e.g., FFR was most sensitive to minimum lumen diameter [10], motivating continued investigation and highlighting the need for mathematically rigorous analyses when assessing the impact of model input uncertainty. Investigations expanded on these initial investigations and integrated UQ strategies in patient-specific frameworks, including studies on the coronary circulation [37], [38], coronary artery bypass grafts [39], and abdominal aorta [40], [41]. Even when evaluating the same vascular territory, it is not straightforward to compare results across studies, largely due to how data are sampled and uncertainty is characterized. However, recognizing that numerous sources of uncertainty, coupled with the high computational costs associated with 3D patient-specific modeling, necessitate the use of efficient approaches to achieve tractable results. Integrating efficient sampling methods [42], surrogate models, and dimensionality-reduction techniques [11], all provide promise to reduce computational burden while preserving the fidelity of UQ in hemodynamic simulations. For example, herein `UncertainSCI` utilized a WAFP sampling approach and PCE techniques, both of which are computationally efficient for characterizing the variability in simulations due to uncertainty in model-dependent quantities. Collectively, these studies and others underscore the growing importance of UQ in hemodynamic simulations and highlight the need for continued development to fully realize its potential in promoting confidence in patient-specific models in the clinical setting.

Input parameter values and associated distributions were obtained from the literature and supplemented with patient-specific data where available (Table 1). While the analytical solution parameter specifications were restricted to literature values, the results provided general insights into WSS under uncertainty. Findings from patient-specific analysis highlighted the need for additional data collection to advance model confidence. For example, uncertainty in viscosity (literature-derived) and velocity (measured) accounted for ~99% of the variability in WSS in the patient-specific model (combined unary and their binary interactions). Unlike routine anticoagulation monitoring in the catheterization lab (e.g., activated clotting time), blood viscosity is rarely quantified. Importantly, however, viscosity values vary with factors such as hematocrit, sex, disease severity, and even altitude [43]. The presented data indicate the importance of quantifying patient-specific viscosities, which could, perhaps with point-of-care devices [44], reduce uncertainty in predicted hemodynamic metrics. Unlike viscosity, velocity data were derived from invasive measurements; however, data were collected in the left main coronary artery across only 6 cardiac cycles. Although velocity variability was low, additional data could reduce measurement variability and uncertainty in model predictions. Collectively, more comprehensive personalized data acquisition and reporting would promote model credibility, and the presented framework is sufficiently modular to incorporate future advances in medical technologies aimed at quantifying hemodynamics metrics with greater fidelity.

Clinical studies have demonstrated that hemodynamic forces are markers of plaque progression and vulnerability[16], [45]. Yet, UQ-certified computational models have not been linked to clinical data or patient outcomes to provide confidence in model predictions. The lack of translation of UQ techniques to guide clinical decision-making motivates efforts to couple uncertainty-aware computational models with clinical observations to establish confidence in model-based predictions. In conjunction with the presented results, which reveal strong sensitivity of WSS values to uncertain inputs, one could reasonably question the credibility of the conclusions drawn in these clinical associations. Accordingly, incorporating UQ into these analyses is essential to distinguish robust relationships from those that may simply reflect instability in the underlying model inputs. Such rigor is fundamental to establishing trustworthy and safe modeling frameworks necessary for the clinical adoption of patient-specific computational models.

Although this study makes a valuable contribution to integrating UQ into hemodynamic simulations, the limitations must be acknowledged. Importantly, these assumptions were introduced mainly to reduce computational costs, particularly when running hundreds of patient-specific models, and to avoid confounding the interpretation of results in this initial investigation. First, velocity was assumed to be steady. This assumption represents a first step toward developing the coupled UQ-FE CFD framework and interpreting results, with future efforts aimed at capturing variability in time-dependent quantities (e.g., unsteady flows) through, for example, Karhunen-Loève expansions [46]. Second, Poiseuille flow assumes the cylinder is rigid, and the patient-specific model likewise assumes a rigid wall. Although solutions for steady flow of a viscous fluid in a thin-walled elastic cylinder exist and fluid-structure interaction modeling techniques are well established [47], [48], integrating material stiffness introduces additional sources of uncertainty that increase study complexity and computational cost. Third, the fluid was considered Newtonian. While constitutive equations characterizing the non-Newtonian viscosity of blood are well developed [49], incorporating the additional viscous terms that arise from the shear rate-dependent viscosity increases model complexity and would require further assumptions on parameter distributions (Fig. 1). Finally, radius was considered an uncertain parameter in the analytical solution, but the geometry was constant in the patient-specific model. In contrast to material property and boundary condition uncertainty, including geometric uncertainty is non-trivial, particularly due to the need for a non-intrusive, automated model construction and discretization pipeline [50], [51].

In conclusion, this study presents a framework for characterizing forward model uncertainty in analytic and CFD simulations of coronary artery hemodynamics. We report that uncertainties in hemodynamic parameters propagate through the governing equations, leading to appreciable variability in WSS values. These findings highlight the need for improved data acquisition to reduce epistemic uncertainties and support future extensions to more complex hemodynamic conditions. Ultimately, this work demonstrates the critical importance of UQ in evaluating the credibility of patient-specific simulations and provides a foundation for developing computational tools that more reliably support clinical decision-making and mechanically-driven mechanistic investigations.

**ACKNOWLEDGMENTS**

Portions of this research were conducted with the advanced computing resources provided by Texas A&M High Performance Research Computing. The authors thank Drs. Habib Samady and David Molony for access to the clinical data.

**FUNDING**

This study was supported by funding from the National Institutes of Health – R01 HL150608 (L.H.T.) and U24 ED-029012 (A.N.).

**Appendix: Parameter Set Oversampling**

PCE convergence was evaluated in the Poiseuille solution for WSS using 2×  oversampling, with the number of parameter sets doubled at each order. That is, parameter sets of 28, 40, 60, 92, and 132 were evaluated at orders 1, 2, 3, 4, and 5, respectively (cf. Table 2).

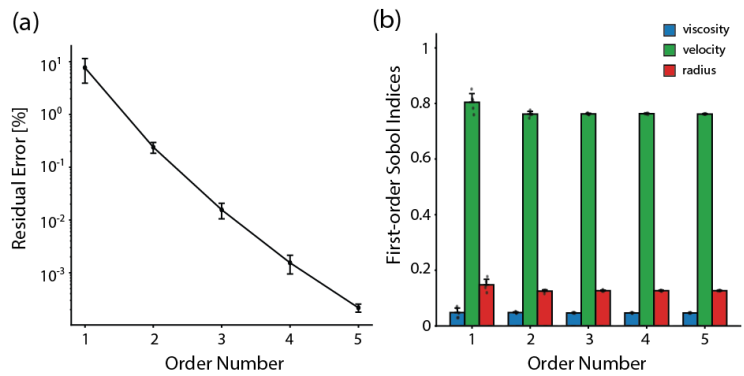

**Fig. A1** Evaluation of PCE emulator and sensitivity indices convergence for the analytical solution - (a) relative error and (b) first-order Sobol indices at 2x oversampling.